\documentclass{IEEEtran}
\usepackage[T1]{fontenc}
\usepackage[latin9]{inputenc}
\usepackage{graphicx}
\begin{document}

\title{Glomerular protein separation as a mechanism for powering renal concentrating
processes}

\author{Robyn F.R. Letts, David M. Rubin, Robert H. Louw, Diane Hildebrandt}
\maketitle
\begin{abstract}
Various models have been proposed to explain the urine concentrating
mechanism in mammals, however uncertainty remains regarding the origin
of the energy required for the production of concentrated urine. We
propose a novel mechanism for concentrating urine. We postulate that
the energy for the concentrating process is derived from the osmotic
potentials generated by the separation of afferent blood into protein-rich
efferent blood and protein-deplete filtrate. These two streams run
in mutual juxtaposition along the length of the nephron and are thus
suitably arranged to provide the osmotic potential to concentrate
the urine. The proposed model is able to qualitatively explain the
production of various urine concentrations under different clinical
conditions. An approach to testing the feasibility of the hypothesis
is proposed. 
\end{abstract}

\section{Introduction}

The renal mechanism for concentrating the urine in mammals is incompletely
understood and is widely believed to involve the presence of molecular
pumps capable of pumping solutes against their concentration gradients
\cite{Sands2008}; a particularly energy-intensive process. Alternative
explanations for the process have been proposed which do not invoke
the concept of membrane pumps \cite{Sands2008}, however these so-called
passive models do not provide an explanation for the origin of the
energy in the concentrating process. Indeed, some authors even argue
against the very existence of these energy-dependent pumps \cite{Ling2001,Pollack2001}.
For a model of the renal concentrating mechanism to be consistent,
it would need to explain the observed urine concentrating phenomena
and correctly predict what would happen under various clinical circumstances.

\section{Background}

The primary functions of the kidney are ridding the body of nitrogenous
metabolic waste products, maintaining an appropriate electrolyte balance,
and ensuring the body's water balance is maintained. This is achieved
through the production of small volumes of concentrated urine during
times of dehydration (anti-diuresis) and large volumes of dilute urine
during times of over hydration (diuresis).

\subsection{Macroscopic measurements}

What is known and measurable at a macroscopic level is that metabolic
waste-laden blood is delivered to the kidneys at mean arterial pressure
via the renal arteries, and slightly ``cleaner'' blood is removed
at central venous pressure via the renal veins. The waste products,
excess solutes and water are removed from the kidneys in the form
of urine. A cortico-medullary concentration gradient exists from the
boundary between the cortex and medulla to the tip of the papilla
and is present in the kidneys of all mammals \cite{Sands2008}, apparently
maximised during anti-diuresis and reduced during diuresis \cite{Maril2005,Hai1969,Knepper1986}.
Glomerular filtration rate (GFR) is the filtration rate of plasma
in the kidney and is approximately 20~\% of renal plasma flow, averaging
about 180~litre/day in healthy humans \cite{Guyton2011}.

\subsection{Microscopic structure}

There are estimated to be between 0.3 and 1.4 million nephrons in
the human kidney \cite{Nyengaard1992}, specifically arranged with
their various structural parts (renal corpuscle, proximal convoluted
tubule, loop of Henle, distal convoluted tubule and collecting duct)
and their closely associated blood supply in either the cortex or
medulla. Peritubular capillaries surround the cortical structures
while the vasa recta surround the medullary structures. This suggests
structural and functional heterogeneity of the kidney \cite{Evans2004}.

\subsection{Urine concentration}

Urine varies in concentration according to the hydration state of
the animal. At the extremes of over-hydration and dehydration, humans,
for example, are said to be able to dilute their urine to approximately
1/6 the concentration of plasma and concentrate their urine to approximately
4 times the concentration of plasma, respectively \cite{Sands2008,Guyton2011}.
The associated volumes of urine vary from approximately 18~litres
to 0.5~litre. Urea (the main nitrogenous waste in mammals) is water
soluble and therefore requires a certain obligatory volume of water
to be excreted along with it, thus limiting the concentrating ability
\cite{Guyton2011}. Vascular flow within the medulla also has the
effect of limiting urine concentrating ability because of its dissipative
effect on the cortico-medullary concentration gradient \cite{Sands2008,Guyton2011}.

\subsubsection{Traditional model}

The traditional model of the urine concentrating process involves
ultrafiltration of plasma at the glomerulus and subsequent reabsorption
and secretion processes (some passive and others active) along the
length of the nephron, ultimately yielding urine \cite{Sands2008}.
The explanation for the establishment of the cortico-medullary gradient
is countercurrent multiplication of a so-called single effect which
is a small osmotic pressure difference between flows in parallel ascending
and descending tubules of the nephron \cite{Kuhn1942}. In the outer
medulla, the single effect is said to involve pumping sodium chloride
out of the ascending limb of the loop of Henle, against its concentration
gradient \cite{Rocha1973,Ullrich1963}, to provide the osmotic gradient
to promote water movement out of the descending limb and collecting
duct and back into the blood \cite{Sands2008}. It has been shown
that the single effect in the inner medulla is not due to sodium pumping
\cite{Morgan1968,Smith1959} and various hypotheses have been put
forward to explain it, but none have been satisfactory \cite{Sands2008}.

\subsubsection{Anti-diuretic hormone}

Differential urine concentration is apparently mediated through the
effect of Arginine Vasopressin or Anti-diuretic Hormone (ADH) on the
kidney \cite{Robertson2008}. During dehydration, high osmolality
of the blood \cite{Robertson2008,Bankir2001} or low plasma volume
\cite{Bankir2001} causes ADH to be released from the posterior lobe
of the pituitary, while during over-hydration, ADH is not released. 

Tubular ADH receptors are located on the walls of the collecting duct
and respond to ADH by increasing the collecting duct permeabilty to
water. Water is then able to pass from the relatively dilute filtrate
into the concentrated interstitium of the medulla and surrounding
vasa recta, thereby being returned to the blood and conserved. In
comparison, during over-hydration, the release of ADH is not stimulated
and therefore the walls of the collecting ducts remain relatively
impermeable to water and the dilute filtrate passes out of the kidney
as dilute urine \cite{Wade1981}.

\subsection{Energy considerations in the concentration of urine}

In terms of fundamental thermodynamic principles, and as pointed out
by others \cite{Sands2008}, to avoid violating the law of conservation
of energy, the energy put into the system must account for the osmotic
work performed in creating concentrated urine. The energy-intensive
molecular pumps needed in the traditional model have been shown to
require more energy than a cell has available to use \cite{Pollack2001,Ling1962,Ling1997},
therefore casting doubt on the thermodynamic integrity of the traditional
model. A number of authors have considered the energy requirements
of urine production \cite{Vilbig2012,Stephenson1974}; however the
source of the energy remains uncertain.

\section{Hypothesis}

In addition to requiring energy intensive molecular pumping of solutes
against their concentration gradients, the traditional model of the
urine concentrating mechanism appears to have ignored the osmotic
effect of protein remaining in the efferent blood. We propose that
the separation of afferent blood into filtrate, which is relatively
protein-free, and efferent blood, which is more concentrated in serum
protein than the afferent blood, by the process of ultrafiltration
at the glomerulus, provides the chemical potential energy for all
the remaining concentrating mechanisms in the tubules. We suggest
that the overall urine concentrating mechanism in mammals may not
require more energy than is provided at the initial separation step
and that there is therefore no need to propose the molecular pumping
of solutes.

\section{Proposed model}

The proposed model for the urine concentrating mechanism in mammals
needs to explain how, in humans for example, 180~litre/day of filtrate
(at almost plasma concentration of 290~mOsm/litre \cite{Guyton2011})
is converted to 0.5~litre/day concentrated urine (at about four times
plasma concentration) in the case of dehydration. The hypothesis must
also be consistent with various known clinico-pathological conditions,
in order to be considered a viable model worthy of further study. 

We propose a model whereby the only energy input into the system is
the energy of protein separation, which occurs during ultrafiltration
at the glomerulus. The energy for ultrafiltration is derived from
the hydrostatic pressure originating in the cardiovascular system.
Osmotic separations constitute real and usable stored energy, and
mixing may be used to recover this energy as work \cite{Bruhn2014,Achilli2010,Logan2012}.
Indeed, industrial-scale osmotic energy storage and harvesting are
being tested \cite{Achilli2010}. 

If the energy of separation at the glomerulus exceeds the energy required
to concentrate urine, then it may be concluded that an external energy
source, such as energy-intensive molecular pumping in the traditional
model, is not needed, and that the proposed model, at a high level,
is thermodynamically viable. 

\begin{figure}
\begin{centering}
\caption{Schematic diagram illustrating a compartment model of the kidney\label{fig:Compartment diagram}}

\par\end{centering}

\centering{}\includegraphics[scale=0.4]{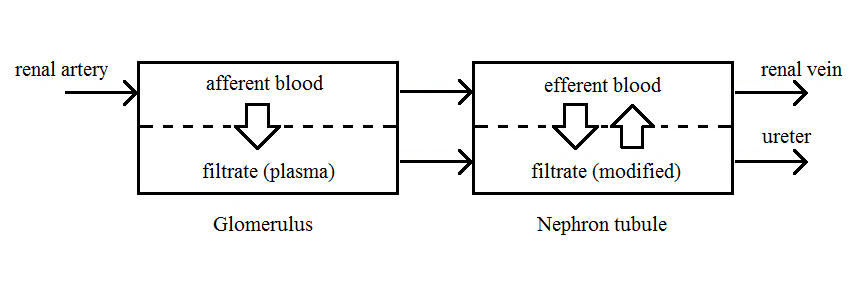}
\end{figure}

A typical nephron has been redrawn as a high level compartment model
in figure \ref{fig:Compartment diagram}.
Afferent blood originating from the renal artery enters the first
blood chamber (i.e. representing the glomerular capillary network
in the Bowman's capsule) and a portion is filtered (approximately
20\% \cite{Guyton2011}) across the membrane due to the net pressure
difference (hydrostatic and oncotic) of approximately 10~mmHg \cite{Guyton2011}.
The resulting filtrate resembles plasma in composition but without
serum proteins \cite{Guyton2011} (i.e. the membrane is permeable
to small solutes but not to larger negatively-charged protein molecules),
while the efferent blood is simply lacking a portion of plasma and
is therefore concentrated (by 20\%) in serum protein. Albumin has
a complex non-linear behaviour in terms of its osmotic coefficient
and when concentrated, it appears to be substantially more osmotically
active \cite{Lin2001,Cameron2006}. The oncotic pressure in the efferent
blood may thus be far greater that the simple 20\% increase in concentration
would imply. The filtrate resides in the filtrate compartment (i.e.
within the nephron tubule lumen), and is able to interact osmotically
with the efferent blood in the second blood chamber (i.e. peritubular
capillaries in the cortex or vasa recta in the medulla), separated
by the nephron membrane that may be differentially permeable to different
solutes. If the energy of separation is sufficient to concentrate
urine, then the separation process can be thought of as a battery
being charged up. The potential energy stored in this osmotic battery
is then drawn down to run the more distal concentrating processes.
Once the filtrate has traversed the length of the nephron, it leaves
its compartment via the ureter as urine, while the efferent blood
is returned to general circulation via the renal vein.

\section{Predictions and Justification}

The normal urine concentrating process can be qualitatively explained
by the model proposed in this paper. During dehydration, the model
predicts concentrated urine production, mediated by ADH secretion.
Tubular receptors change the water permeability of the nephron membrane
(i.e. reducing resistance to water flow from the filtrate compartment
into the efferent blood compartment), causing more water to be reabsorbed
and urine to be concentrated. As long as the energy of separation
exceeds the energy needed to concentrate the urine, changes in membrane
permeability due to the effect of ADH secretion should be sufficient
to explain the increased urine concentration.

The fact that newborn mammals and birds cannot create concentrated
urine \cite{Calcagno1954,Liu2001,Schwartz1992,Winberg1959,Edelmann1960}
is typically explained by immaturity of the organ \cite{Sands2008}.
However, it has also been found that infants have low serum protein
levels relative to adult values \cite{Rosenthal1997,Darrow1933}.
This finding fits with our hypothesis that low serum protein levels
reduce the ability of the kidneys to provide the necessary chemical
potential energy for concentrated urine to be produced. 

A clinical observation that can be explained by the proposed model,
and not by traditional models, is that of renal failure following
hepatic failure. The pathogenesis of this so-called hepato-renal syndrome
is unclear, and there is no visible renal histopathology \cite{Flint1863,Arroyo1996}.
Furthermore, administration of albumin appears to be renoprotective
in this condition \cite{Lee2012} and renal function typically resumes
after successful liver transplantation \cite{Koppel1969,Iwatsuki1973}.
We suggest that our hypothesis explains these findings on the grounds
of hypoalbuminemia accompanying liver failure \cite{Sands2008}. According
to our hypothesis, a reduction in serum protein level would mean a
reduced ability to set up the chemical potential to drive the tubular
processes, resulting in renal failure. Similarly, an increase in serum
protein level following liver transplantation or albumin administration
would provide the chemical potential necessary for urine concentration
and improved kidney function.

Finally, this model would suggest that severe proteinuria may result
in diminished renal function because the presence of protein in the
filtrate represents inadequate separation at the glomeruli, and thus
less osmotic potential for the distal processes. The association between
proteinuria and diminishing renal function has been described but
the mechanism remains unclear \cite{Abbate2006,Thomas2011}.

\section{Conclusion}

The proposed model is simple in its conception and in its current,
qualitative state, it is able to provide a qualitiative explanation
for the observed phenomena in urine concentration and predict what
happens under certain pathological conditions. The next step is to
quantify the model parameters, run simulations to test the model and
ultimately determine whether the hypothesis could be valid. As with
any scientific hypothesis, falsifiability is an important and necessary
feature \cite{Popper1959}. We propose that this can be achieved thorough
a quantitative thermodynamics analysis of the energy balances in the
proposed process.

\bibliographystyle{IEEEtran}
\bibliography{Kidney_database}

\end{document}